\documentclass[aps,prd,showpacs,twocolumn,superscriptaddress,nofootinbib,preprintnumbers]{revtex4-2}

\usepackage{aas_macros}
\usepackage{arydshln}
\usepackage{makecell}

\usepackage{array}
\newcolumntype{C}{|c|}
\usepackage{tabularx}
\usepackage{empheq}

\newcommand{\eV}{\,\text{eV}}

\newcommand{\wa}{w_0w_a{\rm CDM}}
\newcommand{\ld}{\Lambda{\rm CDM}}

\usepackage{float}

\usepackage{amssymb, amsmath, bm, dcolumn, epsf, graphicx, latexsym, slashed} 
\usepackage[utf8]{inputenc}
\usepackage{multirow}

\usepackage{color}

\def\be{\begin{equation}}
\def\ee{\end{equation}}
\def\bea{\begin{eqnarray}}
\def\eea{\end{eqnarray}}

\definecolor{dgreen}{rgb}{0,0.6,0.0}

\usepackage{hyperref}
\usepackage{comment}
\usepackage{booktabs}
\usepackage{diagbox}

\usepackage[normalem]{ulem}

\begin{document}

\title{
Modified gravity constraints with Planck ISW-lensing bispectrum
}

\author{Anton Chudaykin}\email{anton.chudaykin@unige.ch}
\affiliation{D\'epartement de Physique Th\'eorique and Center for Astroparticle Physics,\\
Universit\'e de Gen\`eve, 24 quai Ernest  Ansermet, 1211 Gen\`eve 4, Switzerland}
\author{Martin Kunz}\email{martin.kunz@unige.ch}
\affiliation{D\'epartement de Physique Th\'eorique and Center for Astroparticle Physics,\\
Universit\'e de Gen\`eve, 24 quai Ernest  Ansermet, 1211 Gen\`eve 4, Switzerland}
\author{Julien Carron}\email{julien.carron@unige.ch}
\affiliation{D\'epartement de Physique Th\'eorique and Center for Astroparticle Physics,\\
Universit\'e de Gen\`eve, 24 quai Ernest  Ansermet, 1211 Gen\`eve 4, Switzerland}

\begin{abstract} 
We present updated constraints on modified gravity by including the Integrated Sachs-Wolfe (ISW) effect from CMB lensing–CMB temperature cross-correlations, based on the latest Planck PR4 maps.
Utilizing the Effective Field Theory of dark energy approach and adopting the $w_0w_a$CDM background cosmological model, we find that including the CMB ISW lensing cross-correlations tighten constraints on the modified gravity parameters by approximately $20\%$, reducing the viable parameter space by $40-80\%$.
We derive constraints from Planck CMB, Planck and ACT CMB lensing, DESI DR1 BAO, CMB ISW-lensing, and type Ia supernovae (SN Ia) data.
The constraints on the EFT parameters controlling the kinetic braiding and non-minimal coupling are consistent with General Relativity (GR) at the 95\% CL.
In particular, we obtain a bound on the kinetic braiding parameter, $c_B < 1.2$ at 95\% CL.
In the $w_0$–$w_a$ parameter space, our results imply a crossing of the phantom divide, $w=-1$.
The modified gravity model shows a mild preference over $\ld$ at the $1.8\sigma$, $2.6\sigma$ and $3.2\sigma$ levels for the combinations with Pantheon+, Union3 and DESY5 supernova datasets.
We find that using the latest \texttt{HiLLiPoP}+\texttt{LoLLiPoP} likelihoods alleviates the departure of modified gravity parameters from the GR-values compared to results using {\it Planck} 2018 data.
This paper underlines the importance of the ISW lensing probe in constraining late-time modifications of gravity.
\end{abstract}

\maketitle

\section{Introduction}
\label{sec:intro}

Dark energy, the unknown physical mechanism behind the observed accelerating expansion of the Universe, constitutes one of the major mysteries of fundamental physics. A key question is whether the dark energy is simply a static cosmological constant or whether it is a dynamical mechanism that evolves over time. Recent observations by the Dark Energy Spectroscopic Instrument (DESI) and external data favor to some extent a dynamical dark energy~\cite{DESI:2024mwx}. However, the preferred evolution requires the dark energy equation of state to cross the `phantom divide' where $p=-\rho$. This behavior is difficult to obtain without pathologies in most simple models (e.g.~\cite{Carroll:2003st,Kunz:2006wc,Creminelli:2008wc}) but it can achieved in the Horndeski class of general single-field scalar-tensor theories, for example with kinetic gravity braiding \cite{Deffayet:2010qz}. This motivated multiple studies to analyze the DESI data in the context of general scalar-tensor theories, e.g.~\cite{Chudaykin:2024gol,Ye:2024ywg,Ishak:2024jhs,Wolf:2024stt}.

The effective field theory of dark energy (EFT of DE) approach~\cite{Creminelli:2008wc,Bellini:2014fua} provides a powerful framework to describe general modifications of gravity in a unified manner. 
In this formalism, all the possible modifications of Einstein equations consistent with the symmetries of the cosmological background and gauge freedom are parameterized by a set of fixed operators, multiplied by free time-dependent functions.
In the context of Horndeski theories, the dynamics of linear perturbations associated with a scalar degree of freedom is governed by four arbitrary functions of time, termed $\alpha$-functions.
Constraints on these functions serve as a test for deviations from General Relativity (GR).

Here we revisit the analysis of the general scalar-tensor Horndeski theory of ref.~\cite{Chudaykin:2024gol}. 
In that work, we found that the EFT-based approach is able to fit the data as well or better than the standard, phenomenological model used in \cite{DESI:2024mwx}.
In order to improve constraints and test more rigorously whether modifications of GR are indeed supported by the data, we would like to include additional data sets that are able to specifically probe modified gravity type theories.

Modified gravity theories can in particular change the evolution of metric perturbations at late times, leading to a different integrated Sachs-Wolfe (ISW) effect.
This makes the ISW effect a promising probe of modified gravity.
The ISW is imprinted in the CMB temperature power spectrum on large angular scales.
However, this effect is only $\mathcal{O}(10\%)$ for $\ld$ and is thus swamped by cosmic variance.
The ISW effect can be more efficiently isolated by cross-correlating the CMB with tracers of large-scale structure~\cite{Giannantonio:2008zi,Giannantonio:2012aa,Ho:2008bz} (see ref.~\cite{Planck:2015fcm} for a review).
In this work, we employ the CMB temperature cross-correlations with the CMB lensing potential. 
The late-time ISW lensing cross-correlation signal\footnote{We will in the following refer to the ISW lensing cross-correlation also as the ISW lensing bispectrum~\cite{Lewis:2011fk}, since the CMB lensing signal is obtained with the help of a quadratic estimator.} is detected in the Planck PR4 maps at about a $4\sigma$ level~\cite{Carron:2022eum}, which suggests that it could indeed be a useful tool to probe modified gravity effects.

In this work, we 
first explore the sensitivity of the ISW lensing bispectrum to modified gravity parameters in the EFT of DE approach.
Second, we derive the parameter constraints from the Planck CMB, Planck+ACT CMB lensing, CMB ISW-lensing, DESI DR1 BAO, and SN Ia data.
We present results for three different supernova datasets -- PantheonPlus~\cite{Scolnic:2021amr}, Union3~\cite{Rubin:2023ovl}, and DESY5~\cite{DES:2024tys} -- that were studied in the official DESI analysis~\cite{DESI:2024mwx}.
Third, we compare the results with those obtained in the standard $\wa$ model.

We improve our previous analysis~\cite{Chudaykin:2024gol} in several directions.
First, we employ the most recent \texttt{HiLLiPoP}+\texttt{LoLLiPoP} likelihoods~\cite{Tristram:2020wbi,Tristram:2023haj} based on the Planck PR4 maps.
This CMB dataset is free from the $A_L$ and $\Omega_k$ anomalies, which are known to correlate with modifications of gravity~\cite{Ishak:2024jhs,Specogna:2024euz}, 
allowing us to obtain robust constraints.
Second, we include the ISW lensing bispectrum data, which tightens the constraints on modified gravity parameters. 
Third, we utilize three different SN Ia datasets -- PantheonPlus, Union3, and DESY5 -- presenting results for each separately.
This allows us to explore potential discrepancies between the supernova datasets and separate them from the effects of modified gravity.

The paper is organized as follows. In Sec.~\ref{sec:data}, we describe our analysis methodology and present cosmological datasets used in the analysis. Our final results are presented in Sec.~\ref{sec:res}. We conclude in  Sec.~\ref{sec:conc}. Additional parameter constraints are presented in App.~\ref{app:full}.

\section{Analysis procedure}
\label{sec:data}

We explore the space of general scalar-tensor Horndeski theory within the EFT of DE approach \cite{Gubitosi:2012hu}. Our analysis adopts the formalism of \cite{Bellini:2014fua}, where the perturbation evolution is governed by several $\alpha$-functions, while the background can be chosen independently. 
The time-dependence of the $\alpha$-functions is parameterized as $\alpha_i(a) = c_i \Omega_{\rm DE}(a)$, where $a$ is the scale factor and $c_i$ is a free constant parameter. 
We fix the kinetic term $c_K = 1$ and set the speed of gravitational waves equal to the speed of light, $c_T = 0$, and we allow the braiding parameter $c_B$ as well as the Planck mass run rate parameter $c_M$ to vary.
We parametrize the background evolution through a time-evolving dark energy equation of state, $w(a)=w_0+w_a(1-a)$. 
Additionally, we impose the no-ghost and no-gradient stability conditions, for details see \cite{Chudaykin:2024gol}.
We refer to this modified gravity model as ``MG''.

In addition, we also study the popular $\wa$ scenario where the dark energy corresponds to an effective quintessence model with the same equation of state, $w(a)=w_0+w_a(1-a)$. As such models cannot cross $w=-1$, we use the standard parametrized post-Friedmann approach~\cite{Fang:2008sn} in this case to keep the perturbations finite.

\subsection{Methodology}
\label{sec:data2}

Parameter constraints obtained in this work are obtained with the Einstein-Boltzmann code \texttt{hi\_class}~\cite{Zumalacarregui:2016pph,Bellini:2019syt}, interfaced with the \texttt{Montepython} Monte Carlo sampler~\cite{Audren:2012wb,Brinckmann:2018cvx}. 
\texttt{hi\_class} solves the modified linear perturbation equations for the metric and scalar-field in the synchronous gauge within the EFT of DE framework.
To predict the late-time ISW effect in the CMB, the code computes temperature transfer functions using modified gravitational potentials~\cite{Lesgourgues:2013bra}.

We perform Markov Chain Monte Carlo (MCMC) analyses
to sample the posterior distributions using the Metropolis-Hastings algorithm. We adopt a Gelman-Rubin convergence criterion $R-1<0.03$. The plots and marginalized constraints are generated with the publicly available \texttt{getdist} package.~\footnote{\href{https://getdist.readthedocs.io/en/latest/}{https://getdist.readthedocs.io/en/latest/}} 
The minimum $\chi^2$ values are computed using the
simulated-annealing optimizer \texttt{Procoli} package~\cite{Karwal:2024qpt}.~\footnote{\href{https://github.com/tkarwal/procoli}{https://github.com/tkarwal/procoli}}

In the minimal $\ld$ model we vary 6 standard cosmological parameters: $\omega_{\rm cdm}=\Omega_ch^2$, $\omega_b=\Omega_bh^2$, $H_0$, $\log(10^{10}A_s)$, $n_s$ and $\tau$. In the $\wa$ scenario, we sample two additional parameters, $(w_0,w_a)$, which parameterize a time-dependent equation of state for dark energy. We adopt uniform priors on the dark energy parameters $w_0\in[-3,1]$ and $w_a\in[-3,2]$, following~\cite{DESI:2024mwx}.
In Hormdeski gravity theory, we additionally vary two EFT amplitudes, $c_B$ and $c_M$, by imposing flat uninformative priors $c_i\in[-10,10]$.
In all scenarios, we assume a spatially flat Universe with two massless and one massive neutrino species with $m_\nu=0.06\eV$. 
In the $\ld$ and $\wa$ models we utilize the Halofit non-linear matter power spectrum. Since the Halofit fitting formula was not calibrated against modified gravity scenarios, we conservatively adopt the linear matter power spectrum in the general Horndeski theory. 
This choice has a minimal impact on the final results, shifting the posterior means of the modified gravity parameters by less than $0.2\sigma$~\cite{Chudaykin:2024gol}.

\subsection{Data}
\label{sec:data3}

We use the following datasets in our MCMC analyses:

\subsubsection{CMB}

We employ as our baseline the high-$\ell$ TTTEEE \texttt{HiLLiPoP} and low-$\ell$ EE \texttt{LoLLiPoP} likelihoods from the {\it Planck} PR4 data release~\cite{Tristram:2020wbi,Tristram:2023haj}, supplemented with the low-$\ell$ TT \texttt{Commander} likelihood~\cite{Planck:2018vyg}. 
In addition to the primary CMB, we include measurements of the lensing potential power spectrum, $C_\ell^{\phi\phi}$, from {\it Planck} \texttt{NPIPE} PR4 CMB maps~\cite{Carron:2022eyg}, in combination with lensing measurements from the Atacama Cosmology Telescope (ACT) Data Release 6 (DR6)~\cite{ACT:2023kun}.~\footnote{The likelihood is publicly available at \href{https://github.com/ACTCollaboration/act\_dr6\_lenslike}{https://github.com/ACTCollaboration/act\_dr6\_lenslike}; we use \texttt{actplanck\_baseline} option.}
We refer to these CMB datasets as ``CMB''.

To evaluate the impact of the {\it Planck} PR4 data, 
we also report the constraints using the official {\it Planck} (2018) release \cite{Planck:2018nkj}, which includes the high-$\ell$ \texttt{plik} TTTEEE, low-$\ell$ \texttt{SimAll} EE  and low-$\ell$ \texttt{Commander} TT likelihoods.
This dataset is combined with CMB lensing information from the {\it Planck}+ACT combination and is denoted as ``CMB (PR3).''

An alternative {\it Planck} PR4 \texttt{CamSpec} likelihood is available~\cite{Efstathiou:2019mdh,Rosenberg:2022sdy}. 
The results from \texttt{CamSpec} and \texttt{HiLLiPoP}-\texttt{LoLLiPoP} are fully consistent, with the former showing a slight excess in the lensing effect in the CMB power spectra.
Since this effect is degenerate with modifications of gravity, in our analysis we utilize the \texttt{HiLLiPoP}-\texttt{LoLLiPoP} likelihoods, which provide more conservative constraints on modified gravity.

\subsubsection{ISWL}

We utilize measurements of the ISW lensing bispectrum, $C_\ell^{T\phi}$, in the multipole range $2<\ell<84$ from the {\it Planck} CMB
PR4 maps~\cite{Carron:2022eum}.~\footnote{The ISW lensing likelihood is publically available at \href{https://github.com/carronj/planck\_PR4\_lensing}{https://github.com/carronj/planck\_PR4\_lensing}.} 
This measurement was obtained by cross-correlating the {\it Planck} CMB lensing map from~\cite{Carron:2022eyg}, which was derived from CMB multipoles $\ell>100$ using a quadratic estimator that combines temperature and polarization data, with the large-scale temperature data ($\ell<100$). The ISW lensing likelihood incorporates the expected covariance with the large-scale CMB and lensing spectra.
We label this dataset as ``ISWL''.

\subsubsection{DESI}

We employ the DESI DR1 BAO data as implemented in the official DESI likelihood~\cite{DESI:2024mwx}.~\footnote{The DESI BAO likelihood is publicly available at \href{https://github.com/cosmodesi/desilike}{https://github.com/cosmodesi/desilike}.} 
These measurements are split into 7 redshift bins: one bright galaxy sample (BGS) in the redshift range $0.1<z<0.4$, two luminous red galaxy (LRG) samples in $0.4<z<0.6$ and $0.6<z<0.8$, one combined LRG + emission line galaxy (ELG) sample in $0.8<z<1.1$, one ELG sample in $1.1<z<1.6$, one quasar sample spanning $0.8<z<2.1$ and one Lyman-$\alpha$ Forest sample covering $1.77<z<4.16$. We use the label ``DESI'' for this dataset.

\subsubsection{SN}

We employ three different SN Ia datasets.
First, we utilize the PantheonPlus compilation~\cite{Scolnic:2021amr}, which consists of $1550$ spectroscopically confirmed SN Ia in the refshift range $0.01<z<2.26$.
The second dataset considered is the Union3 sample~\cite{Rubin:2023ovl}, containing $2087$ SN Ia with redshifts $0.01<z<2.26$.
The third SN Ia dataset is the Year 5 supernova analysis from the Dark Energy Survey, which includes 194 low-redshift SN Ia within $0.025<z<0.1$ and 1635 photometrically classified SN Ia spanning $0.1<z<1.3$~\cite{DES:2024tys}.
For each dataset, we analytically marginalize over the absolute magnitude to closely follow the DESI analysis~\cite{DESI:2024mwx}.~\footnote{We made the SN Ia likelihoods publicly available at \href{https://github.com/ksardase/sn\_marg}{https://github.com/ksardase/sn\_marg}.}
We refer to these measurements as ``PantheonPlus'', ``Union3'' and ``DESY5'', respectively.
Additionally, we collectively denote these supernova datasets as ``SN''.

\section{Results and Discussions}
\label{sec:res}

\subsection{ISW lensing bispectrum}
\label{sec:resISW}

In this section, we examine the sensitivity of the ISW lensing bispectrum to modifications of gravity within the EFT of DE framework.

We begin by exploring the dependence of $C_\ell^{T\phi}$ on the choice of EFT parameters.
To that end, we fix the standard 6 cosmological parameters, as well as $w_0$ and $w_a$, to that of the best-fit MG model posterior obtained in the CMB (PR3)+DESI+PantheonPlus analysis of~\cite{Chudaykin:2024gol}.
We vary the remaining EFT parameters $c_M$ and $c_B$, which control the dynamics of linear perturbations associated with a scalar degree of freedom.

\begin{figure}[!t]
\includegraphics[width=1\columnwidth]{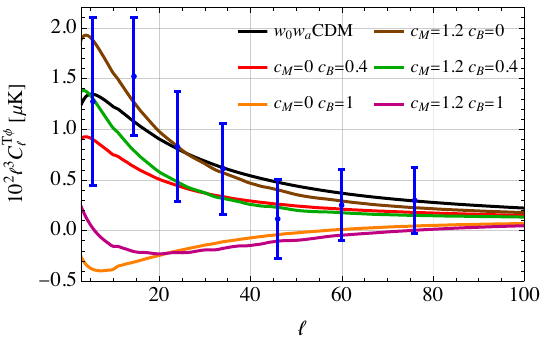}
\caption{
Linear theory predictions for $C_\ell^{T\phi}$ in the MG model while fixing the standard 6 cosmological parameters and $(w_0,w_a)$ to their best-fit values obtained in the CMB (PR3)+DESI+PantheonPlus analysis of ref.~\cite{Chudaykin:2024gol}.
Black line represents the $\wa$ prediction for the same best-fit cosmology, where the dark energy perturbations are computed using the parametrized post-Friedmann approach.
The blue points with $1\sigma$ error bars represent the ISW lensing bispectrum estimate in the range $2<\ell<84$ from the {\it Planck} PR4 maps~\cite{Carron:2022eum}. 
}
\label{fig:CTphi_fix} 
\end{figure}

Fig.~\ref{fig:CTphi_fix} shows the linear theory predictions for $C_\ell^{T\phi}$ for various representative values of $c_M$ and $c_B$. 
Modified gravity models tend to suppress the power of the ISW lensing signal compared to that in $\wa$ ($c_M=c_B=0$).
This result is attributed to moderate kinetic gravity braiding, which reduces the late ISW effect~\cite{Renk:2016olm,Zumalacarregui:2016pph}.
Positive values of the Planck mass run rate counteract this suppression by enhancing $C_\ell^{T\phi}$ on large scales.
This opens up a new degeneracy direction between $c_M$ and $c_B$.
Importantly, the ISW lensing measurement disfavors models with large positive $c_B$.
This highlights the potential of ISW lensing cross-correlations as a probe of modified gravity.

Other cosmological measurements are sensitive to modifications of gravity.
The ISW-induced change to the gravitational potentials affects low multipoles of the CMB TT spectrum. This can also have an impact on the peak of the CMB lensing power spectrum, where it is somewhat sensitive to low redshifts. To determine whether the ISW-lensing signal provides additional information gain, we take into account constraints from the full set of cosmological data.

For this purpose,
we define a set of reference MG models as follows.
We select five distinct values of $c_M$, approximately uniformly distributed in the range $[-0.6,1.2]$.
For each selected $c_M$, we randomly choose points from the MCMC chains obtained in the CMB (PR3)+DESI+PantheonPlus MG analysis of ref.~\cite{Chudaykin:2024gol}, 
ensuring that the $\chi^2$ values deviate by less than $2$ from the best-fit value.
~\footnote{All parameters of the best-fit model lie well within the 95\% confidence intervals of the corresponding posterior distribution.}
This selection guarantees that the reference MG models provide a good description of the cosmological data.
We emphasize that we do not fix any parameters; instead, we sample {\it all} parameters to achieve a good fit to the data.

\begin{figure*}[!t]
\includegraphics[width=1\columnwidth]{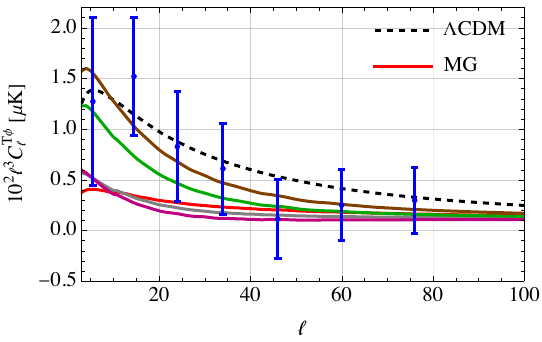}
\includegraphics[width=1\columnwidth]{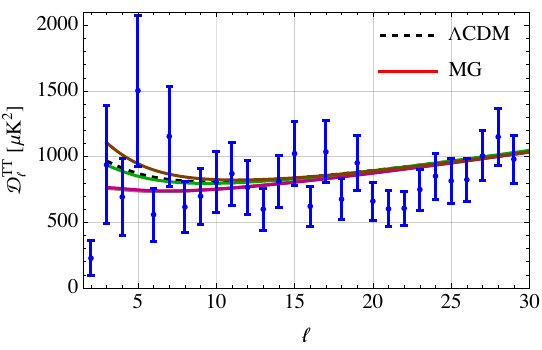}
\caption{
Theoretical predictions for the ISW lensing bispectrum $C_\ell^{T\phi}$ ({\it left panel}) and the temperature power spectrum $C_\ell^{TT}$ ({\it right panel}) in the MG model for different parameter choices selected from the MCMC chains of the CMB (PR3)+DESI+PantheonPlus MG analysis in ref.~\cite{Chudaykin:2024gol} (for details on the selection algorithm, see the main text).
For clarity, we report only the EFT parameters: $(c_M,c_B)=(-0.55,1.02)$ (in red), $(0.14,0.61)$ (in gray, best-fit model), $(0.43,0.69)$ (in purple), $(0.77,0.36)$ (in green), and $(1.20,0.31)$ (in brown).
The $\Lambda$CDM prediction (dashed line) corresponds to the best-fit model from the same dataset.
The blue points with $1\sigma$ error bars represent the $C_\ell^{T\phi}$ estimate from the {\it Planck} PR4 \texttt{NPIPE} maps~\cite{Carron:2022eum} and the $C_\ell^{TT}$ measurement from the \texttt{Commander} likelihood~\cite{Planck:2018vyg}. 
}
\label{fig:CTphi} 
\end{figure*}

Fig.~\ref{fig:CTphi} compares the predicted $C_\ell^{T\phi}$ and $C_\ell^{TT}$ spectra for the randomly chosen reference models with the {\it Planck} measurements. 
For $c_B\lesssim1$, the modified gravity models lead to very similar ISW lensing signals, which lie below the data on large scales. 
In particular, the statistical significance of the difference between the best-fit MG model (in gray) and the data is $2.8\sigma$. 
This illustrates that the ISW lensing bispectrum has an extra sensitivity to modified gravity beyond what is captured by CMB, BAO, and SN Ia data.
For a moderate braiding parameter, the $C_\ell^{T\phi}$ data requires high values of the Planck mass run rate, $c_M\gtrsim1$, while still maintaining a good fit to the low-$\ell$ $C_\ell^{TT}$ data.
This behavior is attributed to the degeneracy direction introduced by the CMB ISW lensing data.

Our results suggest that adding the ISW lensing cross-correlations into the analysis can significantly tighten constraints on the modified gravity parameters. The $C_\ell^{T\phi}$ probe is particularly sensitive to kinetic gravity braiding, by excluding scenarios with $c_B\sim1$.
Additionally, the ISW lensing data necessitates somewhat higher values of $c_M$, which can be important when including additional external probes, e.g. the galaxy-CMB cross-correlations~\cite{Stolzner:2017ged,Seraille:2024beb} and the weak lensing and galaxy clustering data~\cite{DES:2022ccp}. 
The addition of ISW lensing data can also improve the bound on $c_M$ due to the correlation in the $c_M-c_B$ plane.

\subsection{Parameter constraints in MG model}
\label{sec:resMG}

Here, we report results from MCMC analyses exploring the full parameter space. 
Throughout, we quote mean values with 68\% confidence intervals.
The one-dimensional marginalized constraints are summarized in Tab.~\ref{tab:par}.

We start by presenting the results from the combination of CMB and DESI data with the various type Ia supernova datasets. For the EFT coefficients, we obtain
\begin{widetext}
\begin{equation*}
\label{pars}
\mbox{\parbox{0.15\textwidth}{\centering$\rm CMB\!+\!DESI$\\$\rm +PantheonPlus$}}
\left\{
\begin{aligned}
&c_M=1.00_{-1.28}^{+0.68},\\
&c_B=0.72_{-0.45}^{+0.26},
\end{aligned}
\right.  
\quad
\mbox{\parbox{0.13\textwidth}{\centering$\rm CMB\!+\!DESI$\\$\rm +Union3$}}
\left\{
\begin{aligned}
&c_M=0.60_{-1.18}^{+0.50},\\
&c_B=0.73_{-0.39}^{+0.25},
\end{aligned}
\right. 
\quad
\mbox{\parbox{0.13\textwidth}{\centering$\rm CMB\!+\!DESI$\\$\rm +DESY5$}}
\left\{
\begin{aligned}
&c_M=0.66_{-1.13}^{+0.62},\\
&c_B=0.71_{-0.40}^{+0.25}.
\end{aligned}
\right. 
\end{equation*}
\end{widetext}
These results are consistent with no-running of the Planck mass ($c_M=0$).
The data indicate a mild preference for positive values of $c_B$.
In particular, the $c_B$ posterior deviates from zero at the $2\sigma$, $2.5\sigma$ and $2.4\sigma$ significance levels for the combinations with PantheonPlus, Union3 and DESY5 respectively.

The $\Delta\chi^2_{\rm min}$ values between the maximum a posteriori of the MG model and the maximum a posteriori of the concordance cosmology are $-9.3$, $-13.9$ and $-17.9$ for the combinations of CMB and DESI with PantheonPlus, Union3 and DESY5.
These correspond to preferences for MG over $\ld$ at the significance levels of $1.9\sigma$, $2.7\sigma$ and $3.2\sigma$, respectively. 
These preferences are somewhat weaker compared to those observed for the $\wa$ scenario relative to $\ld$~\cite{DESI:2024mwx,Chudaykin:2024gol}.

\begin{figure*}[!t]
\includegraphics[width=0.325\textwidth]{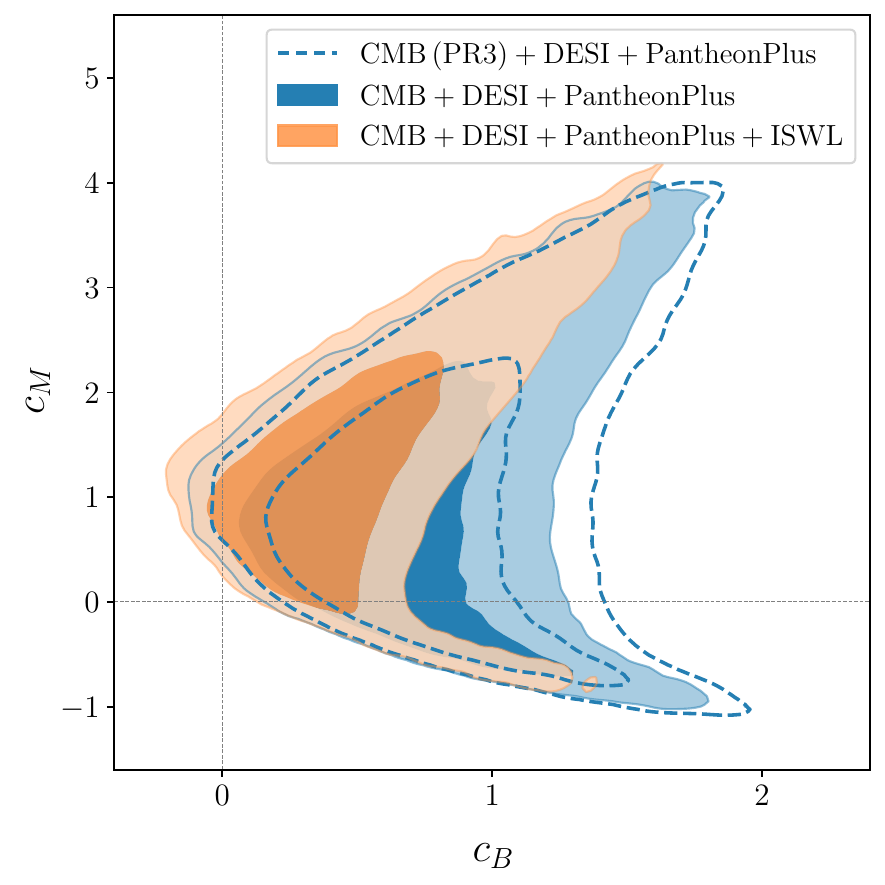}
\includegraphics[width=0.325\textwidth]{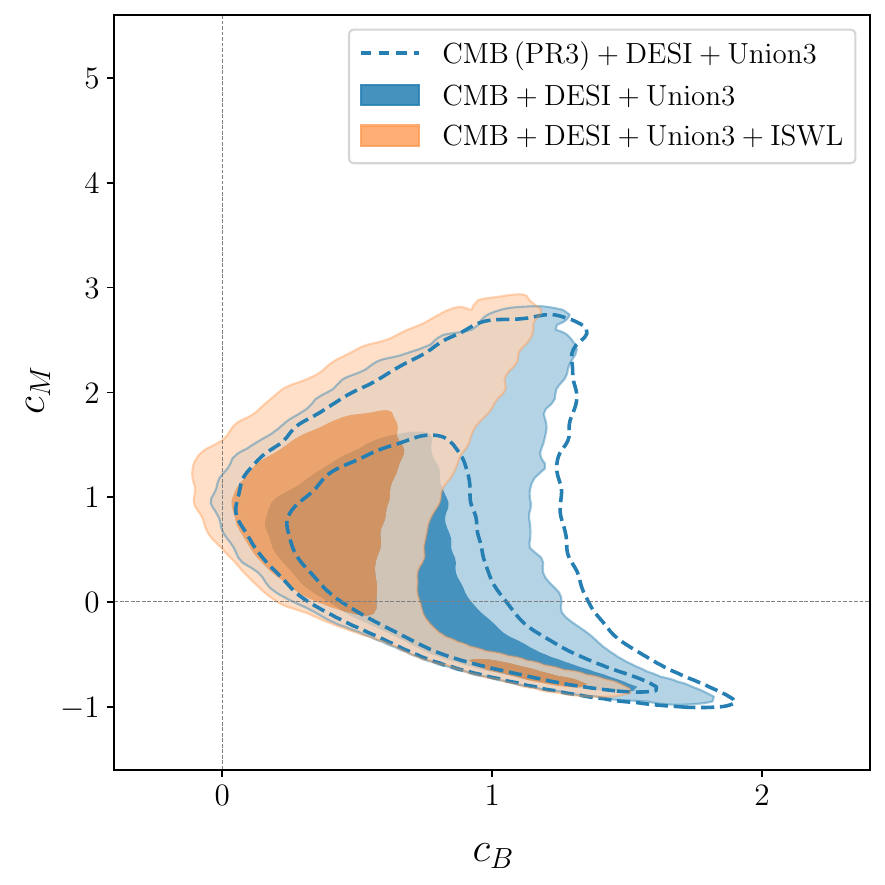}
\includegraphics[width=0.325\textwidth]{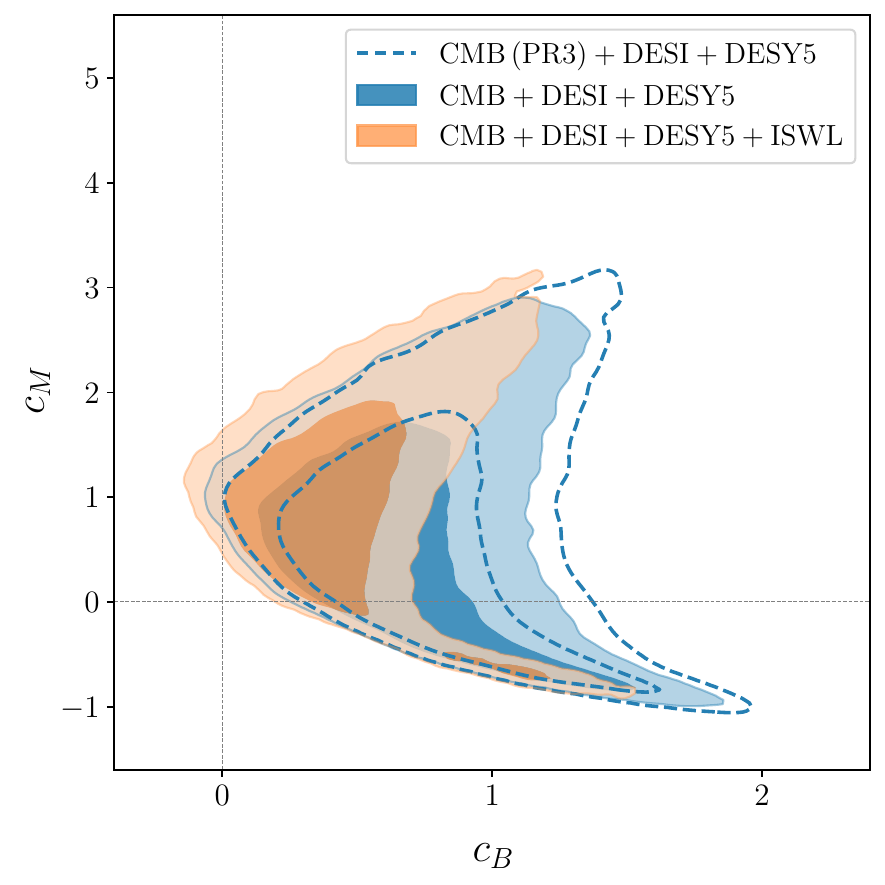}
  \caption{
68\% and 95\% marginalized constraints in the $c_B-c_M$ plane for various data combinations in the MG model. 
Adding the ISW lensing data shrinks the credible-interval contours, excluding scenarios with $c_B\gtrsim1$.
}
\label{fig:cBcM_SN} 
\end{figure*}

Fig.~\ref{fig:cBcM_SN} shows the posterior distributions in the plane $c_B-c_M$.
When the ISWL is not included, the shape of the contours is primarily driven by the late ISW effect in the CMB TT power spectrum~\cite{Zumalacarregui:2016pph}. 
Specifically, the effect of modified gravity conspires to prevent a large power in the CMB TT power spectrum, resulting in the observed degeneracy direction between $c_B$ and $c_M$ parameters.
The lower boundary for $c_M$ is driven by the onset of gradient instabilities~\cite{Ishak:2024jhs}. 

It is important to compare our results with those obtained using {\it Planck} PR3 data. Fig.~\ref{fig:cBcM_SN} illustrates that the posteriors in the $c_B-c_M$ plane shift towards the GR values ($c_B=c_M=0$) when using \texttt{HiLLiPoP}-\texttt{LoLLiPoP} likelihoods based on {\it Planck} PR4 maps.
This indicates that the statistical significance of deviations from GR decreases when moving from {\it Planck} PR3 to {\it Planck} PR4 data.
Specifically, the preference for a non-zero $c_B$ increases to the $2.4\sigma$, $2.8\sigma$, and $2.8\sigma$ levels when using CMB (PR3), DESI with PantheonPlus, Union3, and DESY5, respectively. 
Our results agree with the previous studies~\cite{Ishak:2024jhs,Specogna:2024euz}, which showed that the departure of modified gravity parameters from their GR values is somewhat alleviated when using {\it Planck} PR4 data.

When we add the CMB ISW-lensing data, we find
\begin{widetext}
	\begin{center}
\begin{empheq}[box=\fbox]{equation*}
\label{parsISW}
\mbox{\parbox{0.17\textwidth}{\centering$\rm CMB\!+\!DESI\!+\!ISWL$\\$\rm +PantheonPlus$}}
\left\{
\begin{aligned}
&c_M=1.28_{-1.05}^{+0.63},\\ 
&c_B=0.49_{-0.41}^{+0.17}, 
\end{aligned}
\right.  
\quad
\mbox{\parbox{0.17\textwidth}{\centering$\rm CMB\!+\!DESI\!+\!ISWL$\\$\rm +Union3$}}
\left\{
\begin{aligned}
&c_M=0.93_{-0.77}^{+0.64},\\
&c_B=0.48_{-0.33}^{+0.14}, 
\end{aligned}
\right. 
\quad
\mbox{\parbox{0.17\textwidth}{\centering$\rm CMB\!+\!DESI\!+\!ISWL$\\$\rm +DESY5$}}
\left\{
\begin{aligned}
&c_M=1.01_{-0.82}^{+0.63},\,\,\\
&c_B=0.47_{-0.34}^{+0.14}.\, 
\end{aligned}
\right. 
\end{empheq}
\end{center}
\end{widetext}
First, the addition of the ISW lensing cross-correlations reduces the uncertainty on $c_B$ by approximately $20\%$, while shifting the mean values closer to the GR prediction.
The braiding parameter is now consistent with zero at the 95\% CL across all supernova datasets.\footnote{We note however that models crossing $w=-1$ cannot have $c_B = c_M = 0$ in the EFT of DE framework.}
In particular, we constrain $c_B < 1.2$ at the 95\% CL, independent of the choice of supernova dataset.
Second, the errorbars on $c_M$ decrease by $10\%-20\%$ with the posterior distribution becoming more symmetric when adding the ISW lensing data.
These findings showcase the power of the $C_\ell^{T\phi}$ probe.

Adding the ISWL data also increases the values of the Planck mass run rate by approximately $30\%$.
This effect can be attributed to two main factors.
First, the ISW-lensing bispectrum introduces a new degeneracy direction, which results in slightly higher values of $c_M$.
As discussed in Sec.~\ref{sec:resISW}, higher values of $c_M$ increase the power of $C_\ell^{T\phi}$ on large scales, which provides a better fit to the ISW lensing data in scenarios with a moderate braiding parameter, $c_B\sim0.4$.
Second, non-minimal coupling contributes significantly to ensuring a stable phantom crossing scenario~\cite{Ye:2024ywg,Chakraborty:2025syu}.
Since the mean value of $c_B$ decreases upon adding the ISWL data, and the $w_0$, $w_a$ constraints continue to support phantom crossing behavior, slightly larger values of $c_M$ may be necessary to stabilize dark energy perturbations.

The $\Delta\chi^2_{\rm min}$ values between the MG and $\ld$ models are $-8.8$, $-13.3$ and $-17.4$ for the combinations of CMB, DESI, ISWL with PantheonPlus, Union3 and DESY5 datasets.
These imply very similar preferences of the MG scenario over $\ld$ at the $1.8\sigma$, $2.6\sigma$ and $3.2\sigma$ significance levels, respectively.
Despite the significant improvement in posterior constraints, the difference in maximum a posteriori values remains largely unchanged when incorporating the CMB ISW lensing bispectrum.
This indicates that the MG scenario still has enough freedom to fit the $C_\ell^{T\phi}$ data.
We also provide the best-fit values of the EFT parameters: $(c_M,c_B)=(0.58,0.22)$, $(0.40,0.28)$, and $(0.61,0.19)$ for the combinations with Pantheon+, Union3, and DESY5 supernova datasets, respectively. In all cases, the best-fit parameters lie within the 68\% credible intervals of the corresponding posterior distributions, indicating that prior effects are weak.
The same conclusion holds for the other cosmological parameters.

Fig.~\ref{fig:cBcM_SN} illustrates the impact of the ISW lensing cross-correlations in the $c_B-c_M$ plane for three different SN Ia samples.
Adding the $C_\ell^{T\phi}$ measurement helps exclude a significant region with $c_B\sim 1$ in agreement with the results of Sec.~\ref{sec:resISW}.
The Figure of Merit (FoM)~\cite{Albrecht:2006um}~\footnote{The Figure of Merit (FoM) is defined as the inverse area of the posterior contours, ${\rm FoM}=|{\rm det} \mathbf{C}|^{-1/2}$, where $\mathbf{C}$ is the projected covariance matrix in the given parameter space. A higher FoM value corresponds to a more statistically significant constraint of the model parameters.} in the $c_M-c_B$ parameter space increases upon adding the ISWL data by factors of $1.5$ for PantheonPlus, $1.3$ for Union3, and $1.1$ for DESY5.
The impact of the ISW lensing cross-correlations becomes more pronounced in the full modified gravity parameter space, reducing the volume of the $(w_0,w_a,c_B,c_M)$ space by $80\%$ (PantheonPlus), $60\%$ (Union3) and $40\%$ (DESY5).
This highlights a substantial information gain provided by the ISW-lensing signal.

Our results are consistent with those of ref.~\cite{Seraille:2024beb}, which employ the galaxy-CMB temperature cross-correlations as a probe of the late ISW effect. 
Working within the EFT of DE framework, they find that including ISW constraints rules out the region $c_B\gtrsim1$.
The analysis reports somewhat tighter constraints of modified gravity parameters which can likely be attributed to their assumption of a $\ld$ cosmological background.
We plan to explore the impact of galaxy-CMB cross-correlations in future work. 

We quantify the performance for the MG scenario relative to $\ld$ using a series of metrics.  
Tab.~\ref{tab:Bays} summarize our results.
\begin{table}[!t]
    \renewcommand{\arraystretch}{1.2}
    \centering
    \begin{tabular}{cccc} \midrule  \midrule 
         \multirow{1}{*}{Metric} 
         & PantheonPlus
         & Union3
         & DESY5 \\
         \midrule  
         $\Delta\chi^2_{\rm min}$  
         & $-8.8$ 
         & $-13.3$ 
         & $-17.4$ 
         \\
         $\Delta {\rm AIC}$  
         & $-0.8$ 
         & $-5.3$ 
         & $-9.4$ 
         \\
         $\Delta {\rm DIC}$ 
         & $-1.8$ 
         & $-6.0$ 
         & $-9.7$  
         \\
         \midrule \midrule 
    \end{tabular}
    \caption{The best-fit $\Delta \chi^2$, $\Delta{\rm AIC}$ and $\Delta{\rm DIC}$ values computed for the MG model relative to $\ld$ for the combinations of CMB, DESI, ISWL with PantehonPlus, Union3 and DESY5. 
    For clarity, we use the tags PantheonPlus, Union3, and DESY5 in dataset names.
    }
\label{tab:Bays}
\end{table}
We apply the Akaike Information Criteria (AIC)~\cite{1100705}, defined by ${\rm AIC}=\chi^2_{\rm min}+2N_p$, where $N_p$ represents the number of free parameters of the model.
This criterion sets a penalty proportional to the number of extra parameters introduced by a more complex model.
We also employ the Deviance Information Criterion (DIC)~\cite{Liddle:2007fy}, given by ${\rm DIC}=\chi^2(\theta_p)+2p_D$, where $\theta_p$ denotes the parameters at the maximum a posteriori point, and $p_D=\langle\chi^2\rangle-\chi^2(\theta_p)$ is the effective number of parameters, where the average is over the posterior~\cite{Raveri:2018wln}.
Physically, the DIC measures the improvement of the likelihood, within the region of support of the prior, relative to the number of effective parameters that the data constrain.

Our results suggests no significant preference for the MG scenario when using PantheonPlus, and a moderate preference for MG is found for Union3 and DESY5, according to the Jeffreys’ scale~\cite{Spiegelhalter:2002yvw}. 
The AIC and DIC give similar results, indicating that the posteriors are constrained by the data and not by the priors.  
Among the considered SN Ia datasets, DESY5 exhibits the strongest, though not decisive, preference for the modified gravity scenario.
Recently, ref.~\cite{Efstathiou:2024xcq} raised concerns about potential systematics in the low redshift sample of DESY5 observations; however, ref.~\cite{DES:2025tir} explains the causes for this and maintains the validity of the DESY5 results.

\subsection{Comparison with $\wa$ model}
\label{sec:resw0wa}

In this section, we compare the results obtained in the MG scenario with those derived under the $\wa$ model.  
While both models feature the same description of the cosmological background, the MG scenario offers a self-consistent treatment of linear dark energy perturbations within the EFT approach.
The 1D marginalized constraints are summarized in Tab.~\ref{tab:par}.

\begin{figure*}[!t]
\includegraphics[width=0.325\textwidth]{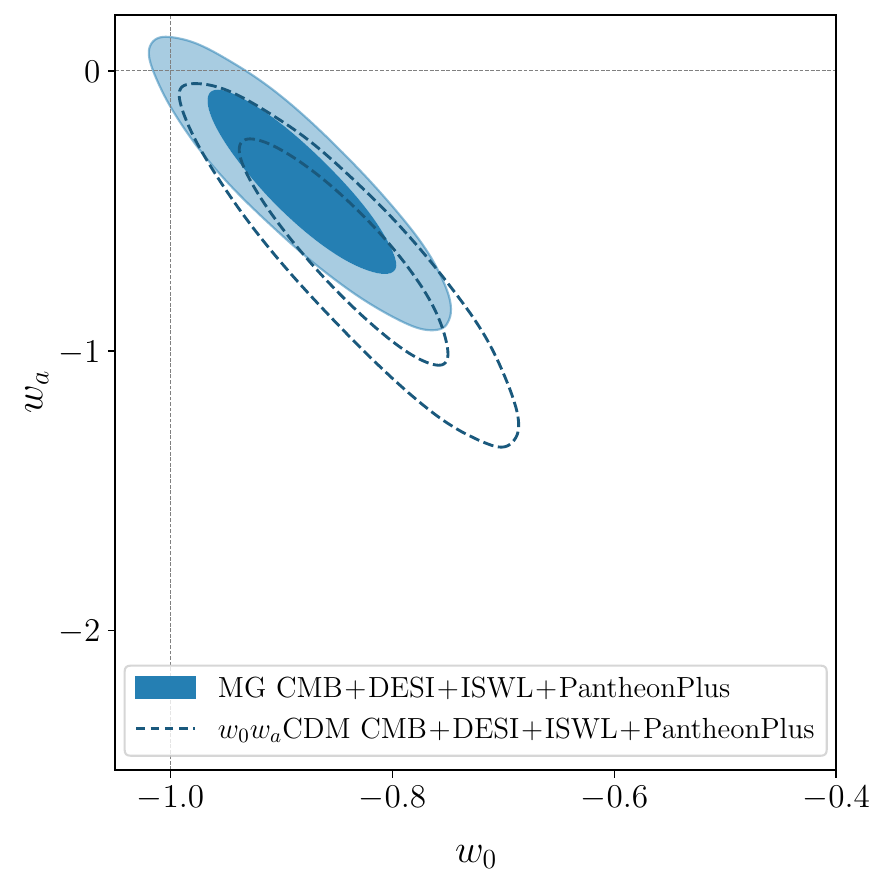}
\includegraphics[width=0.325\textwidth]{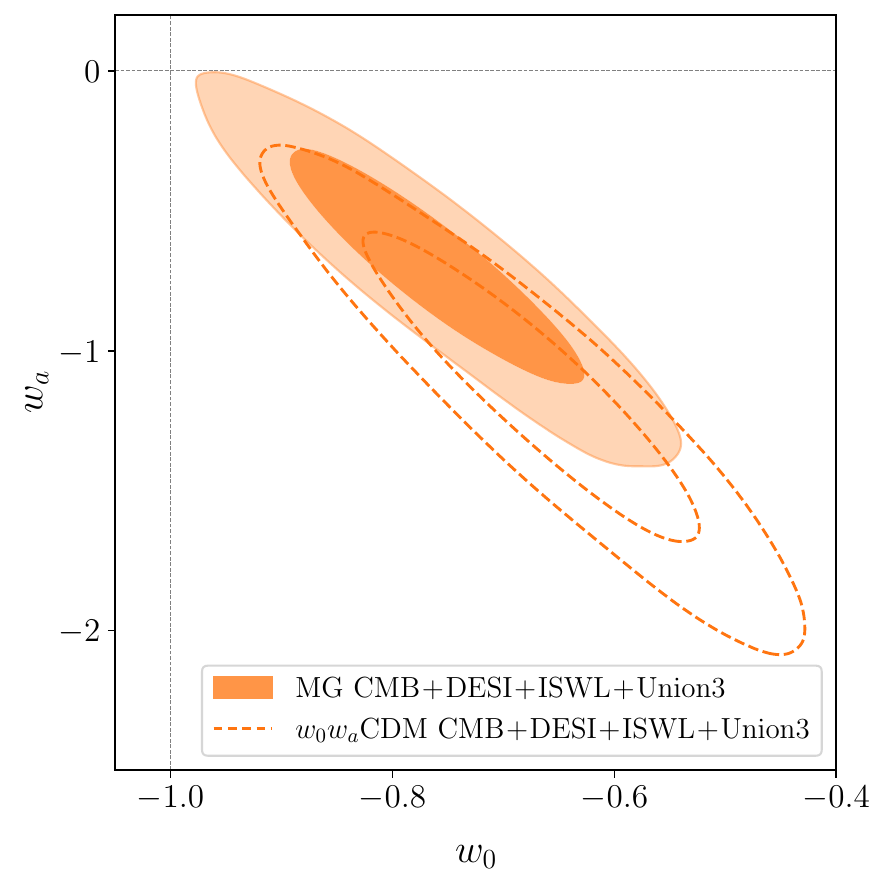}
\includegraphics[width=0.325\textwidth]{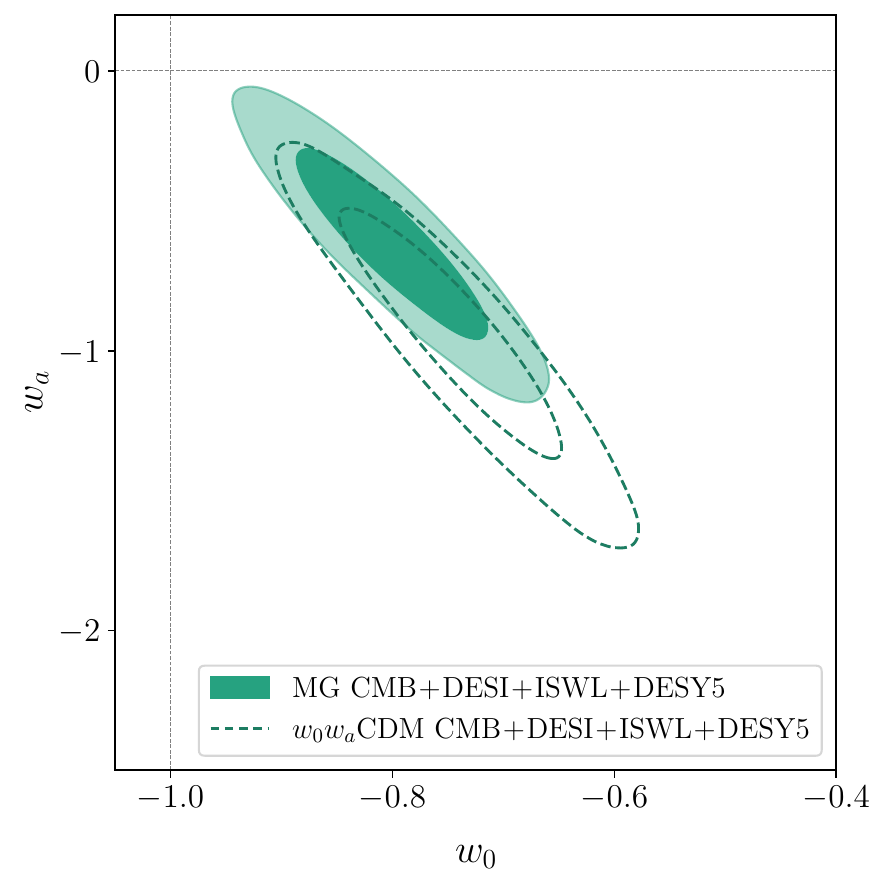}
\caption{
68\% and 95\% marginalized constraints on $w_0$ and $w_a$ parameters in the MG (solid) and $\wa$ (dashed) models.
The constraints are obtained from the combinations of $\rm CMB+DESI+ISWL$ with PantheonPlus (in blue), Union3 (in orange), and DESY5 (in green).
}
\label{fig:w0wa_SN} 
\end{figure*}

Fig.~\ref{fig:w0wa_SN} shows the posterior distributions in the $w_0-w_a$ plane for the MG and $\wa$ scenarios. 
We found that the results are consistent between the two cases, however the $w_0$ and $w_a$ posteriors in the MG model are shifted closer to $\ld$. 
In particular, in the MG scenario the $w_0$ posterior deviates from $-1$  at the significance levels of $2.1\sigma$ (PantheonPlus), $2.8\sigma$ (Union3) and $3.6\sigma$ (DESY5), while in the $\wa$ scenario these shifts increase to $2.7\sigma$, $3.5\sigma$ and $4.1\sigma$, respectively.
In all cases, the MG model favors $w_0>-1$ and $w_a<0$, indicating a phantom crossing at some point in the past.

We now examine the goodness-of-fit for different datasets in the MG and $\wa$ models relative to $\ld$. Tab.~\ref{tab:chi2} presents the breakdown of the best-fit $\Delta\chi^2$ values for each experiment in the CMB+DESI+ISWL+SN analyses.
\begin{table*}[!t]
    \renewcommand{\arraystretch}{1.2}
    \centering
    \begin{tabular}{ccccccc} \midrule  \midrule 
         \multirow{2}{*}{Data} 
         & \multicolumn{2}{c}{PantehonPlus} 
         & \multicolumn{2}{c}{Union3} 
         & \multicolumn{2}{c}{DESY5} \\
         & $\Delta \chi^2_{\rm MG}$ 
         & $\Delta \chi^2_{w_0w_a}$ 
         & $\Delta \chi^2_{\rm MG}$ 
         & $\Delta \chi^2_{w_0w_a}$ 
         & $\Delta \chi^2_{\rm MG}$ 
         & $\Delta \chi^2_{w_0w_a}$ \\
         \midrule  
         CMB primary  
         & $-2.3$ & $-1.1$ 
         & $-2.3$ & $-1.2$ 
         & $-1.8$ & $-0.7$ \\
         CMB lensing & 
         $-1.1$ & $-0.5$ 
         & $-1.0$ & $-0.5$ 
         & $-1.1$ & $-0.6$ \\
         ISWL 
         & $0.0$ & $-0.2$ 
         & $0.0$ & $-0.2$ 
         & $-0.1$ & $-0.2$ \\
         DESI 
         & $-3.0$ & $-3.1$ 
         & $-3.7$ & $-3.7$ 
         & $-4.4$ & $-4.4$ \\
         SN 
         & $-2.3$ & $-2.0$ 
         & $-6.3$ & $-6.3$ 
         & $-10.0$ & $-10.0$ \\
         \midrule 
         Total 
         & $-8.8$ & $-6.9$ 
         & $-13.3$ & $-12.1$ 
         & $-17.4$ & $-15.9$ \\
         \midrule \midrule 
    \end{tabular}
    \caption{$\Delta \chi^2_{\rm model} \equiv \chi^2_{\rm model}-\chi^2_{\Lambda\rm CDM}$ values for the different best-fit models optimized to the CMB+DESI+ISWL+SN datasets, where SN represents PantheonPlus, Union3, and DESY5.
    For clarity, we use the supernova dataset names in dataset labels.
    We split the CMB data into CMB primary (\texttt{LoLLiPoP}+\texttt{HiLLiPoP}+\texttt{Commander}) and CMB lensing ($C^{\phi\phi}_\ell$). 
    }
\label{tab:chi2}
\end{table*}
For this comparison, we evaluate the best-fit models by using the Halofit non-linear matter power spectrum~\cite{Bird:2011rb}.
In the MG scenario, using the linear matter power spectrum underestimates the $C^{\phi\phi}_\ell$ at small angular scales, which leads to a poorer fit to the CMB lensing.
To provide a meaningful comparison, we utilize the Halofit matter power spectrum across all models in this analysis.

Modified gravity provides a better fit to the DESI data, mainly due to an improved description of the BAO signal from the LRG samples at $z_{\rm eff}=0.51$ and $0.71$~\cite{Chudaykin:2024gol,Sapone:2024ltl}.
The MG scenario also demonstrates a slightly better fit to the primary CMB data. 
Interestingly, this improvement increases when using the {\it Planck} 2018 likelihood. For example, in the CMB (PR3)+DESI+ISWL+PantheonPlus analysis the improvement associated with CMB primary data rises to $\Delta\chi^2_{\rm MG}=-5.2$ and $\Delta\chi^2_{w_0w_a}=-2.5$.~\footnote{The combinations with Union3 and DESY5 give similar results.}
Our findings demonstrate that this improvement observed in the MG and $\wa$ scenarios is attributed to a systematic effect in {\it Planck} PR3 data rather than any new physics.
The improvement is not statistically significant when using CMB PR4 data.
This underscores the importance of the {\it Planck} PR4 maps when analyzing modified gravity.

\begin{table*}
\centering
\resizebox{\textwidth}{!}{
\footnotesize
\setcellgapes{3pt}\makegapedcells   
\renewcommand{\arraystretch}{1.3}
\begin{tabular}{lccccccc}
\toprule
\midrule
Model/Dataset & $H_0$ & $\Omega_{\mathrm{m}}$ & $\sigma_8$ & $w_0$ & $w_a$ & $c_M$ & $c_B$\\
\midrule\midrule
\textbf{$\boldsymbol{w_0w_a}$CDM} & & & & & &\\
    \makecell[l]{CMB+DESI+ISWL\\ \quad +PantheonPlus} 
    & $67.89_{-0.71}^{+0.72}$
    & $0.308_{-0.007}^{+0.007}$
    & $0.816_{-0.009}^{+0.009}$
    & $-0.839_{-0.062}^{+0.062}$
    & $-0.66_{-0.24}^{+0.29}$
    & -- 
    & -- \\
    \makecell[l]{CMB+DESI+ISWL\\ \quad +Union3}
    & $66.44_{-0.94}^{+0.94}$
    & $0.322_{-0.010}^{+0.010}$
    & $0.805_{-0.010}^{+0.010}$
    & $-0.668_{-0.100}^{+0.100}$
    & $-1.15_{-0.35}^{+0.38}$
    & -- 
    & -- \\
    \makecell[l]{CMB+DESI+ISWL\\ \quad +DESY5}
    & $67.13_{-0.65}^{+0.65}$
    & $0.315_{-0.006}^{+0.006}$
    & $0.811_{-0.009}^{+0.009}$
    & $-0.742_{-0.066}^{+0.066}$
    & $-0.95_{-0.27}^{+0.32}$
    & -- 
    & -- \\
\midrule
\textbf{MG} & & & & &\\
\makecell[l]{CMB (PR3)+DESI\\ \quad +PantheonPlus} 
    & $67.86_{-0.71}^{+0.71}$
    & $0.308_{-0.007}^{+0.007}$
    & $0.838_{-0.022}^{+0.015}$
    & $-0.866_{-0.061}^{+0.061}$
    & $-0.50_{-0.26}^{+0.26}$
    & $0.96_{-1.33}^{+0.73}$
    & $0.84_{-0.46}^{+0.29}$
    \\
    \makecell[l]{CMB (PR3)+DESI\\ \quad +Union3}
    & $66.54_{-0.94}^{+0.95}$
    & $0.321_{-0.009}^{+0.009}$
    & $0.822_{-0.019}^{+0.014}$
    & $-0.719_{-0.103}^{+0.092}$
    & $-0.91_{-0.32}^{+0.39}$
    & $0.55_{-1.07}^{+0.57}$
    & $0.80_{-0.38}^{+0.26}$
    \\
    \makecell[l]{CMB (PR3)+DESI\\ \quad +DESY5}
    & $67.07_{-0.66}^{+0.65}$
    & $0.315_{-0.007}^{+0.007}$
    & $0.828_{-0.018}^{+0.013}$
    & $-0.774_{-0.068}^{+0.063}$
    & $-0.76_{-0.26}^{+0.30}$
    & $0.66_{-1.14}^{+0.64}$
    & $0.80_{-0.41}^{+0.26}$
    \\
\cdashline{1-8} \noalign{\vskip 0.25em}
    \makecell[l]{CMB+DESI\\ \quad +PantheonPlus} 
    & $67.69_{-0.70}^{+0.70}$
    & $0.308_{-0.007}^{+0.007}$
    & $0.838_{-0.022}^{+0.015}$
    & $-0.875_{-0.060}^{+0.059}$
    & $-0.44_{-0.23}^{+0.26}$
    & $1.00_{-1.28}^{+0.68}$
    & $0.72_{-0.45}^{+0.26}$
    \\
    \makecell[l]{CMB+DESI\\ \quad +Union3}
    & $66.42_{-0.94}^{+0.95}$
    & $0.320_{-0.010}^{+0.009}$
    & $0.822_{-0.020}^{+0.013}$
    & $-0.731_{-0.101}^{+0.092}$
    & $-0.84_{-0.30}^{+0.37}$
    & $0.60_{-1.18}^{+0.50}$
    & $0.73_{-0.39}^{+0.25}$
    \\
    \makecell[l]{CMB+DESI\\ \quad +DESY5}
    & $66.95_{-0.65}^{+0.65}$
    & $0.315_{-0.006}^{+0.006}$
    & $0.827_{-0.018}^{+0.013}$
    & $-0.784_{-0.067}^{+0.060}$
    & $-0.71_{-0.24}^{+0.30}$
    & $0.66_{-1.13}^{+0.62}$
    & $0.71_{-0.40}^{+0.25}$
    \\
\cdashline{1-8} \noalign{\vskip 0.25em}
\makecell[l]{CMB+DESI+ISWL\\ \quad +PantheonPlus} 
    & $67.63_{-0.69}^{+0.70}$
    & $0.309_{-0.007}^{+0.007}$
    & $0.841_{-0.019}^{+0.016}$
    & $-0.882_{-0.056}^{+0.056}$
    & $-0.40_{-0.21}^{+0.22}$
    & $1.28_{-1.05}^{+0.63}$
    & $0.49_{-0.41}^{+0.17}$
    \\
    \makecell[l]{CMB+DESI+ISWL\\ \quad +Union3}
    & $66.38_{-0.95}^{+0.96}$
    & $0.320_{-0.009}^{+0.009}$
    & $0.826_{-0.018}^{+0.015}$
    & $-0.758_{-0.088}^{+0.088}$
    & $-0.72_{-0.28}^{+0.28}$
    & $0.93_{-0.77}^{+0.64}$
    & $0.48_{-0.33}^{+0.14}$
    \\
    \makecell[l]{CMB+DESI+ISWL\\ \quad +DESY5}
    & $66.86_{-0.65}^{+0.65}$
    & $0.316_{-0.007}^{+0.007}$
    & $0.831_{-0.016}^{+0.014}$
    & $-0.800_{-0.058}^{+0.057}$
    & $-0.62_{-0.22}^{+0.23}$
    & $1.01_{-0.82}^{+0.63}$
    & $0.47_{-0.34}^{+0.14}$
    \\
\midrule\bottomrule
\end{tabular}
}
\caption{Mean and 68\% confidence intervals on the relevant cosmological parameters. $H_0$ is measured in units of $\rm km\,s^{-1}\,Mpc^{-1}$. The main results of this work are presented in the last subsection.}
\label{tab:par}
\end{table*}

\section{Conclusions}
\label{sec:conc}

In this work, we explore the use of the ISW lensing bispectrum data to improve constraints on general Horndeski scalar-tensor gravity theories.
Utilizing the EFT of DE approach and assuming a $\wa$ cosmological background, we derive parameter constraints from CMB, DESI BAO DR1, ISW-lensing and SN Ia datasets.

We find that adding the CMB ISW lensing data tightens constraints on modified gravity parameters by approximately $20\%$, reducing the credible-region areas in the $(w_0,w_a,c_B,c_M)$ space by $40-80\%$, depending on the choice of supernova dataset. 
The inclusion of ISW lensing cross-correlations also results in constraints that are more consistent with $\ld$.
For example, the braiding parameter, $c_B$, is now consistent with zero at the 95\% CL for all dataset combinations, whereas the previous analyses~\cite{Chudaykin:2024gol,Ishak:2024jhs}, which did not include the ISW lensing probe, showed the $\gtrsim2\sigma$ preference for $c_B\neq0$. 
These results highlight the temperature-lensing cross-correlation as a powerful probe for testing modifications of gravity.

In the EFT of DE framework, the constraints in the $w_0$-$w_a$ plane shift closer to the $\Lambda$CDM expectation compared to that in the $\wa$ scenario.
While both models feature the same description of the cosmological background, the EFT of DE framework offers a physical model for dark energy perturbations, which allows a stable crossing of the phantom divide, $w=-1$.

We find that the statistical significance of deviations from GR decreases when moving from {\it Planck} PR3 to the latest \texttt{HiLLiPoP}+\texttt{LoLLiPoP} likelihoods based on the {\it Planck} PR4 maps.
Our findings confirm the results of previous studies~\cite{Ishak:2024jhs,Specogna:2024euz}, which showed that deviations from GR /$\ld$ are alleviated when using {\it Planck} PR4 data.
This highlights the importance of the {\it Planck} PR4 data when constraining modified gravity.

Our analysis can be improved in the following directions. 
First, we can include the full-shape measurements of the power spectrum from BOSS/eBOSS~\cite{eBOSS:2020yzd} or DESI~\cite{DESI:2024hhd}, which tightens constraints on the running of the Planck mass $c_M$~\cite{Ishak:2024jhs}.
Second, our constraints can be enhanced by adding the weak lensing and galaxy clustering data from the $3 \times 2$pt Dark Energy Survey (DES) analysis~\cite{DES:2022ccp}.
Third, cross-correlating the CMB temperature with alternative low-redshifts tracers can provide additional insights into modified gravity~\cite{Stolzner:2017ged,Seraille:2024beb}.
We leave these research directions for future exploration.

\vspace{1cm}
\section*{Acknowledgments}
We thank Nils Schöneberg and Julien Lesgourgues for providing the Planck PR4 likelihood. 
AC thanks Maria Berti for useful discussions.
AC and MK acknowledge funding from the Swiss National Science Foundation.
JC acknowledges support from a SNSF Eccellenza Professorial Fellowship (No. 186879).
Numerical calculations have been performed with the Helios cluster at the
Institute for Advanced Study, Princeton. 

\appendix 

\section{Full Parameter Constraints}
\label{app:full}

\begin{figure*}[ht]
	\begin{center}
		\includegraphics[width=1\textwidth]{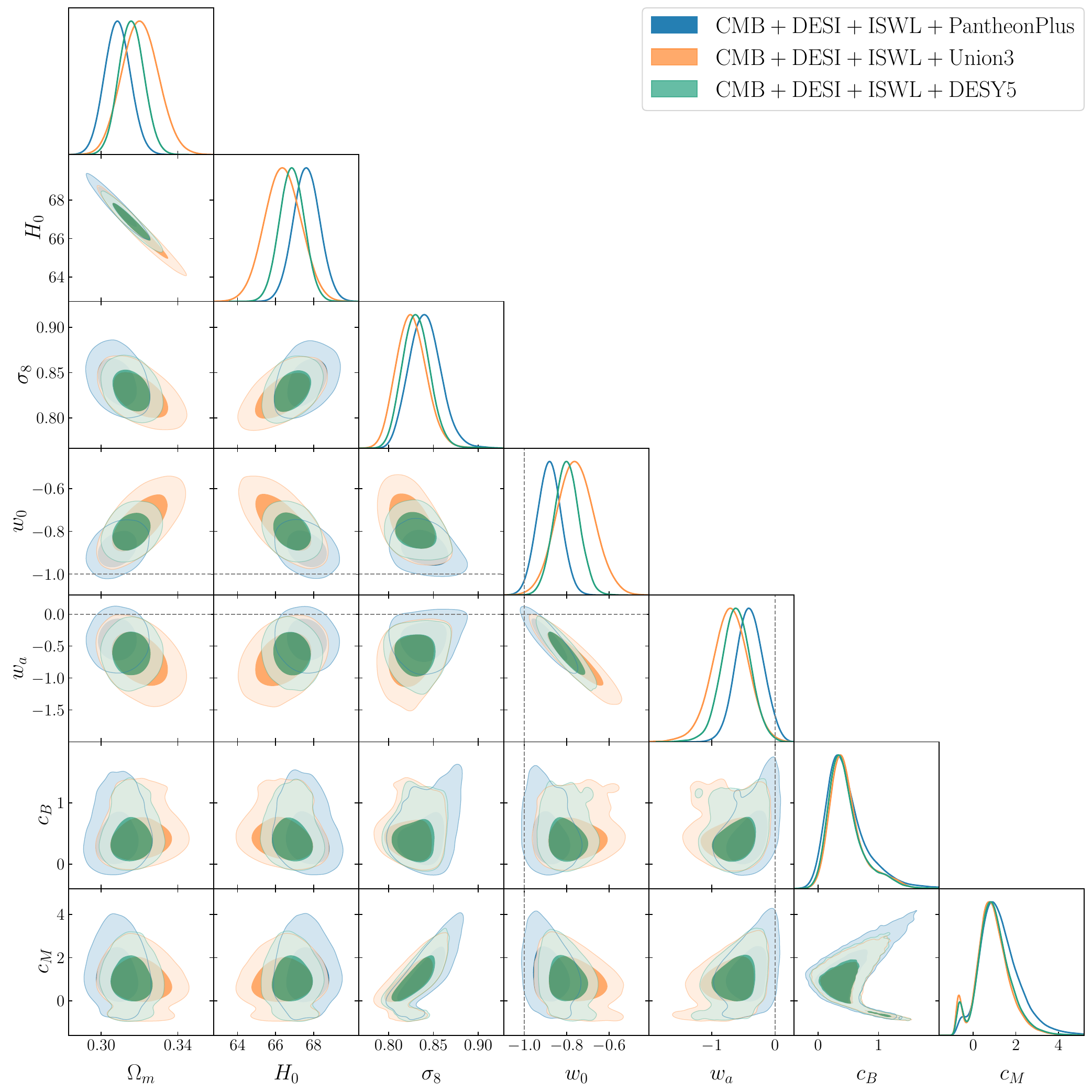}
	\end{center}
	\caption{
		Posterior distributions of the cosmological parameters in the MG model from the combinations of CMB+DESI+ISWL with PantheonPlus (blue), Union3 (orange) and DESY5 (green) datasets. 
		\label{fig:MGfull} } 
\end{figure*}

In this Appendix, we present the posterior distributions for various parameters obtained from the CMB+DESI+ISWL along with three different SN Ia datasets. The results are shown in Fig.~\ref{fig:MGfull}.
The corresponding parameter constraints are listed in Tab.~\ref{tab:par}.

\bibliographystyle{JHEP}
\bibliography{short.bib}

\end{document}